\documentclass[aps,pra,twocolumn,superscriptaddress,showpacs,floatfix]{revtex4-1}
\usepackage{amsmath,amssymb,amsfonts,xspace,epsfig,float,multirow}
\usepackage{graphicx}
\usepackage{amsfonts}
\usepackage[caption=false]{subfig}
\usepackage[export]{adjustbox}

\usepackage[utf8]{inputenc}
\usepackage[english]{babel}

\usepackage{hyperref}
\usepackage{xcolor}
\newcommand{\cs}{}
\newcommand*\dd{\mathop{}\!\mathrm{d}}
\begin{document}

\title{Structural transitions of nearly second order in  classical dipolar gases}
\author{Florian Cartarius}
\affiliation{Universit\'e Grenoble Alpes, Laboratoire de Physique et Mod\'elisation des Milieux Condens\'es, F-38000 Grenoble, France}
\affiliation{CNRS, Laboratoire de Physique et Mod\'elisation des Milieux Condens\'es, F-38000 Grenoble, France}
\affiliation{Theoretische Physik, Universit\"at des Saarlandes, D 66123 Saarbr\"ucken, Germany}
\author{Giovanna Morigi}
\affiliation{Theoretische Physik, Universit\"at des Saarlandes, D 66123 
Saarbr\"ucken, Germany}
\author{Anna Minguzzi}
\affiliation{Universit\'e Grenoble Alpes, Laboratoire de Physique et Mod\'elisation des Milieux Condens\'es, F-38000 Grenoble, France}
\affiliation{CNRS, Laboratoire de Physique et Mod\'elisation des Milieux Condens\'es, F-38000 Grenoble, France}

\date{\today}

\begin{abstract}
Particles with repulsive power-law interactions undergo a transition from a single to a double chain (zigzag) by decreasing the confinement in the transverse direction.  We theoretically characterize this transition when the particles are classical dipoles, polarized perpendicularly to the plane in which the motion occurs, and argue that this transition is of first order, even though weakly. The nature of the transition is determined by the coupling between transverse and axial modes of the chain and contrasts with the behaviour found in Coulomb systems, where the linear-zigzag transition is continuous and belongs to the universality class of the ferromagnetic transition. Our results hold for classical systems with power-law interactions $1/r^\alpha$ when $\alpha > 2$, and show that structural transitions in dipolar systems and Rydberg atoms can offer the testbed for simulating the critical behaviour of magnets with lattice coupling.
\end{abstract}

\maketitle

\section{Introduction}

Strongly-correlated ensembles of ultracold atoms provide an unique platform for simulating dynamics and models predicted for condensed-phase systems, statistical mechanics, as well as to test quantum-field theoretical hypotheses \cite{Lewenstein:07,Bloch:08,Georgescu:14}. Self-organized phases of trapped ions, atoms, and dipolar systems play in this context a prominent role, as they allow one to study and simulate Wigner crystallization \cite{Dubin:99,Astrakharchik:07,Buechler:07}, supersolidity \cite{Goral:02}, and quantum magnetism \cite{Porras:04,Friedenauer:08,Islam:11}, to mention a few examples.

One peculiar instance is the linear-zigzag instability in ion chains. This instability is observed in a linear array of trapped ions by lowering the transverse confinement: Below a critical value the equilibrium configuration is a double array, forming a zigzag chain \cite{Birkl:92}. The transition is continuous and is classically described by a Landau model \cite{Fishman:08}. In the quantum regime,  it is a quantum phase transition of the same universality class of the ferromagnetic transition of an Ising chain in a transverse field \cite{Shimshoni:11,Silvi:13}. The spin order is here associated to the transverse displacement of the ions from the chain axis. It thus naturally offers a testbed for studying, amongst others, kink formation after quenches across the structural transition \cite{kinks} and the spin-Peierls instability \cite{Bermudez:12}. Deep in the quantum regime, where the quantum statistical properties are relevant such as in quantum wires, the linear-zigzag instability is characterized by a rich phase diagram  \cite{Meyer:07}. 

\begin{figure*}
 \subfloat[$\omega_t = 1.1  \ \omega_t^{(c)}$]{
 \includegraphics[width=0.3\textwidth]{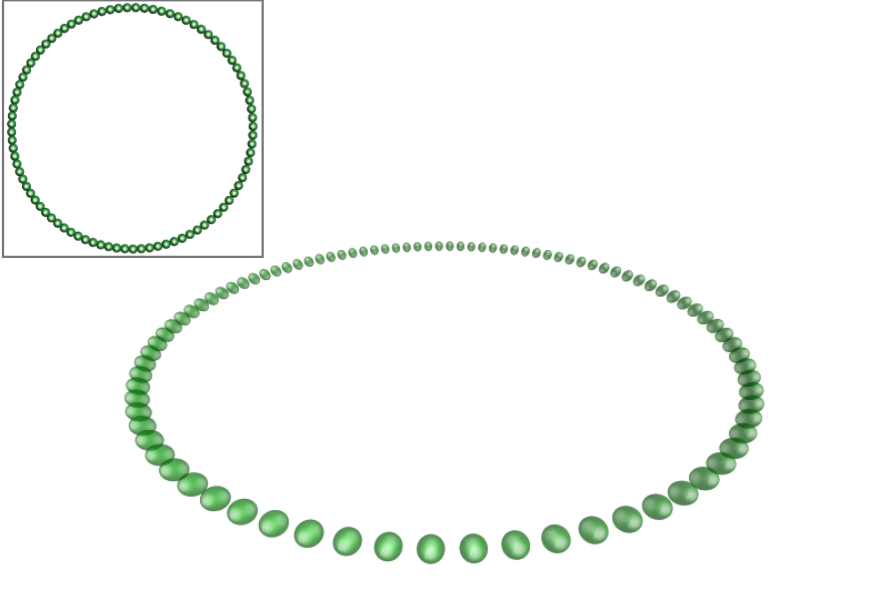}}
 \subfloat[$\omega_t = 0.99  \ \omega_t^{(c)}$]{
  \includegraphics[width=0.3\textwidth]{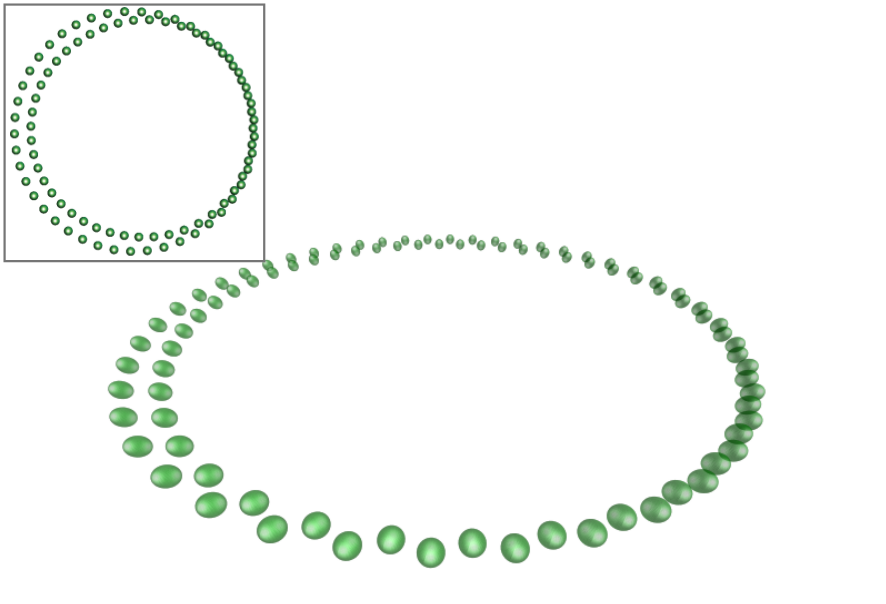}}
  \subfloat[$\omega_t = 0.8  \ \omega_t^{(c)}$]{
 \includegraphics[width=0.3\textwidth]{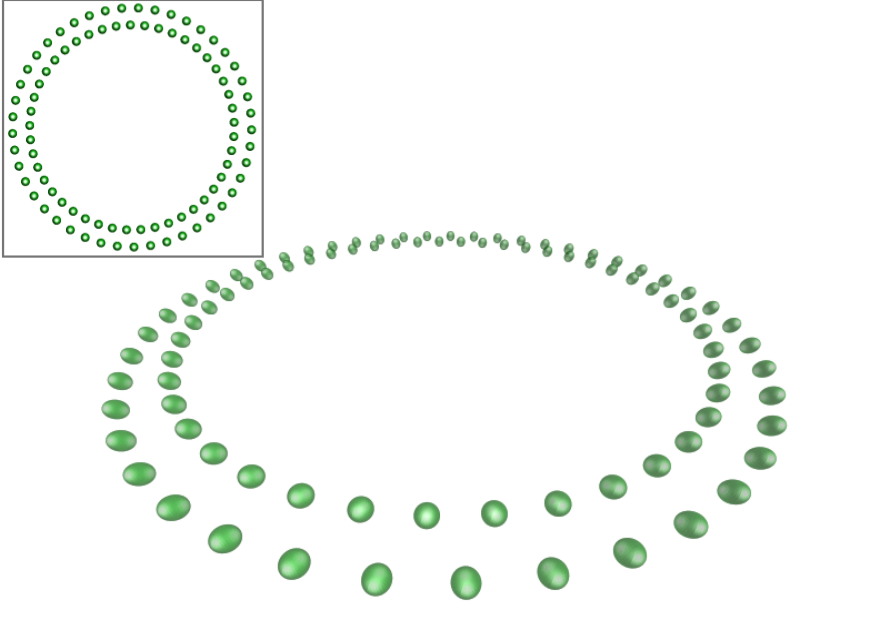}}
 \caption{(Color online) \cs Side view (main panels) and top view (insets) of the various configurations found in the Monte-Carlo simulations: single ring (linear chain) (a), inhomogeneous configuration (b), and double ring (zigzag chain) (c) of classical dipolar particles confined in the plane perpendicular to the polarizing electric field. The different configurations correspond to three decreasing values of the radial confinement in the ring trap. The inhomogeneous configurations as in (b) indicate a coexistence of linear and zigzag structures, and are numerically found using periodic boundary conditions. Similar  structures are found as well in a box with hard walls by varying the transverse frequency or the linear density.  \label{fig:1}}
\end{figure*}


 In this work we analyse linear-zigzag instability in other systems exhibiting repulsive power-law interactions of the type $1/r^\alpha$, \cs focusing in particular on the case $\alpha=3$ corresponding to  dipolar gases. For exponent $\alpha>2$ we show that, in absence of external potentials imposing long-range order, the instability becomes of first order due to the coupling between transverse and axial vibrations, which modifies the critical properties. Quite remarkably, this longitudinal-transverse  coupling among the modes  plays an analogous role as the coupling between spins and phonons for ferromagnetic transitions in compressible lattices \cite{Larkin,Imry}. Evidence  for a first-order transition is brought forward by  the numerical  observation of inhomogenous configurations, indicating that at the instability the chain alternates regions in which the ions exhibit either zigzag or linear order, as shown in Fig. \ref{fig:1}. The regions are separated by kinks whose form is reminiscent of soliton excitations. Such configurations were not reported in previous numerical studies, which analysed the instability for small samples~\cite{Astrakharchik2008,Astrakharchik2009} (composed of about 16 or less dipolar particles), and are observed when the particles number exceeds several tens of particles. Further insight on the nature of the transition  is gained by means of a low-energy theory, which shows that the parameter range in which the inhomogeneous configurations are found shrinks in the thermodynamic limit, even though it remains finite. The transition therefore can be considered as  "weakly" first-order or nearly second order, using the therminology of Refs. \cite{Larkin,Yurtseven:06}. 

This article is organized as follows. In Sec.~\ref{Sec:2} we describe the model and discuss the stability of the ring chain. Monte-Carlo results are presented  Sec.~\ref{Sec:3}. In Sec.~\ref{res:low-energy} we compare the numerical results  with the analytical predictions of the low-energy theory.  Sec.~\ref{res:low-energy} also contains the analysis of the nature of the transition and our predictions for the thermodynamic-limit behaviour.  Finally, Sec.~\ref{Sec:5} discusses the role of thermal fluctuations and offers our concluding remarks.

\section{Physical system}
\label{Sec:2}

We consider $N$ classical particles of mass $m$ which are confined by an anisotropic trap on the $x-y$ plane, assuming a very tight confinement along the $z$ direction. The particles interact via a power-law repulsive potential of the form
\begin{equation}
 V_{\rm int}({\bf r}_1,\ldots, {\bf r}_N) =  \frac{C_D}{2} \sum_{j \neq l}\frac{1}{|{\bf r}_j-{\bf r}_l|^{\alpha}}\,, 
\label{eq:basic_model}
\end{equation}
where $C_D$ is the interaction strength and  ${\bf r}_j=(x_j,y_j)$ is the position of particle $j=1,\ldots,N$. The generic power-law exponent $\alpha$ describes, for instance, the dipolar interaction for $\alpha=3$ (when the particles possess permanent dipoles and are polarized by an external field orthogonal to the plane), or Van-der-Waals interactions for $\alpha=6$. Moreover, the particles are confined by a ring trap of radius $R_0$, which generates the (radially harmonic) potential
\begin{align}
 V_{\rm trap}({\bf r}_1,\ldots, {\bf r}_N) = \frac{1}{2} m \omega_t^2 \sum_{j=1}^N\left( r_j - R_0 \right)^2\,,
\end{align}
with $r_j=|{\bf r}_j|$ and $\omega_t$ the frequency in the radial direction. Such trapping potential is currently realized for quantum gases \cite{Exp,Exp1,Exp2,Exp3,Exp4,Exp7}.
For large radii it approaches a linear trap with periodic boundary conditions.  

We will numerically seek \cs in Sec.\ref{Sec:3} for the configuration which minimizes the energy in the total potential 
\begin{equation}
\label{V:tot}
V=V_{\rm int}+V_{\rm trap}\,,
\end{equation}
close to the linear-zigzag instability.  The regime of stability of the linear configuration is analytically identified by means of a Taylor expansion of the potential about the linear array. This has been performed in Refs. \cite{Astrakharchik2008,Astrakharchik2009}. Below we report the basic steps, here applied to the specific configuration of a ring trap. 

\subsection{Taylor expansion about the equilibrium configuration}

In order to analyse the stability properties of the ring chain, we first rewrite the interaction potential $V_{\rm int}$, Eq. \eqref{eq:basic_model}, in terms of polar coordinates, such that 
$V_{\rm int}=(1/2)\sum_{j,l\neq j}U(r_j,\phi_j,r_l,\phi_l)$.  We then use the center-of-mass and relative coordinates $R_{jl} = (r_j + r_l)/2$, $\rho_{jl} = r_j - r_l$ and $\phi_{jl} = \phi_j - \phi_l$, and cast $U(r_j,\phi_j,r_l,\phi_l)$ into the form 
\begin{align}
 U\left(R_{jl},\rho_{jl},\phi_{jl}\right) = \frac{C_D}{[ \rho_{jl}^2 \cos^2(\phi_{jl}/2) + 4 R_{jl}^2 \sin^2(\phi_{jl}/2) ]^{\frac{\alpha}{2}}}\,.
\label{interactions}
\end{align}
We then perform a systematic expansion of the interaction energy about the configuration in which the ions form a single ring. We denote by $R$ the ring radius, which results to be $R>R_0$ due to the interparticle repulsion.  Moreover, we denote by $a$ the uniform interparticle distance along the ring, such that $a=2\pi R/N$. Assuming that one dipole of the ring is pinned, the single ring is a regular structure which exhibits discrete translational invariance where the particles are located at radial position $r_j=R$ and at angles $\phi_j= 2\pi j/N$ ($j=0,\ldots,N-1$). This configuration corresponds to equilibrium since the first derivatives of the total potential $V$, Eq. \eqref{V:tot}, vanish. In order to verify that the equilibrium is stable, we consider the further terms in the Taylor expansion. Setting  $r_j = R + a \Psi_j$ and $\phi_j = 2 \pi j /N + a \Theta_j / R$, the expansion reads
\begin{widetext}
\begin{align}
 V_{\text{int}} =  \frac 1 2 \sum_{j=1}^N \sum_{l \neq j}^N \sum_{0 \leq n_1 + n_2 + n_3 \leq {\color{black} 6}} \frac{1}{n_1 ! n_2 ! n_3 !}  \frac{a^{n_1+n_2+n_3}}{2^{n_1}R^{n_3}}\frac{\partial^n U\left(R, 0, \phi_j^{(0)} - \phi_l^{(0)}\right)}{\partial R^{n_1} \partial \rho^{n_2} \partial \phi^{n_3}}  (\Psi_j + \Psi_l)^{n_1}(\Psi_j - \Psi_l)^{n_2}(\Theta_j - \Theta_l)^{n_3}, \label{eq:expansion}
\end{align}
\end{widetext}
where $n_1$, $n_2$, $n_3$ are positive integers. In these derivatives all even-order derivatives in $\rho$ vanish because of the symmetry of the single-ring configuration. 

\subsection{Stability of the single ring}

The stability of the linear chain is determined by analysing the Hessian of the second-order derivatives. An analytical expression of the dispersion relation is found using the Fourier modes $\Psi_k$ and $\Theta_k$, such that $ \Psi_j = \frac{1}{\sqrt N} \sum_{k}  \tilde{\Psi}_k e^{i k j a } \label{eq:lin-modes-psi}$,  $\Theta_j =\frac{1}{\sqrt N} \sum_{k} \tilde{\Theta}_k e^{i k j a}$ with  $k=-\pi N/L,\ldots, N\pi/L$ and $L=2\pi R=Na$. Denoting by $V^{(2)}$ the term of the second-order Taylor expansion for $V_{\rm int}$, it takes the form $ V^{(2)} =\sum_k V^{(2)}_k$ with
\begin{align}
 V^{(2)}_k =a^2& \sum_{l \neq 0} \left[ \left|  \tilde\Psi_k \right|^2 \frac{ 1}{ 4} \frac{\partial^2 U(R,0,2\pi l/N)}{\partial R^2} \cos^2\left( k l a /2 \right)  \right.  \nonumber \\
 &+ \left|  \tilde\Psi_k \right|^2  \frac{\partial^2 U(R,0,2\pi l/N)}{\partial \rho^2} \sin^2\left( k l a /2 \right) \nonumber \\
 &+ \left|  \tilde\Theta_k \right|^2 \frac{1}{R^2} \frac{\partial^2 U(R,0,2\pi l/N)}{\partial \phi^2} \sin^2\left( k l a /2 \right)\nonumber \\
 &\left. +     \tilde\Theta_{k} \tilde\Psi_{-k}\frac{1}{4R} \frac{\partial^2 U(R,0,2\pi l/N)}{\partial \phi \partial R}\sin\left(k l a \right)  \right] \,. 
 \label{eq:energy-FT}
\end{align}
For $R,N\to\infty$, but keeping $a=2\pi R/N$ constant, the derivatives with respect to $R$ vanish, such that axial and transverse Fourier modes become decoupled \cite{Astrakharchik2009}.  In this thermodynamic limit, the linear chain is mechanically unstable at $\omega_t=\omega_t^{(c)}(N)$, with $$\lim_{N\to\infty}\omega_t^{(c)}(N)=\sqrt{(93\zeta(5)/8)C_D/(ma^5)}$$
and $\zeta(5)$ the Riemann's zeta function. At this value of the transverse trap frequency the frequency of the transverse mode with quasi momentum $k_0=\pi/a$,  $ \tilde\Psi_{k_0}=\sum_j(-1)^j\Psi_j/\sqrt{N}$, vanishes. The details of the corresponding calculation are reported in Ref. \cite{Astrakharchik2009}. For the Coulomb interaction this instability is a second-order phase transition which is classically described by the Landau model \cite{Fishman:08}. The mode at $k_0$ is then the soft mode driving the instability, and the order parameter the displacement $a\Psi_j$ in the radial direction. In Ref. \cite{Astrakharchik2008,Shimshoni:11,Piacente:10,Altman:12} it has been conjectured that this may hold for any power-law repulsive interaction with $\alpha\ge 1$.  

\section{Minimal-energy configurations}
\label{Sec:3}

We first numerically study the  linear-zigzag instability, focusing on the case $\alpha=3$ of dipolar interactions. We search for the particle configuration which minimizes the total potential energy $V =V_{\rm trap} + V_{\rm int}$ for  different values of the trap frequency $\omega_t$. We  determine the classical ground state of a dipolar \cs  gas  using the Basin-Hopping Monte-Carlo method \cite{MC}, with which we identify the equilibrium configurations corresponding to the global minimum of the potential energy for $N$ ranging from $16$ to $1100$. We note that the configurations we find are expected to reproduce the correct ground state at $T=0$ when the interaction energy exceeds the kinetic energy, hence at sufficiently high densities and for large permanent dipoles \cite{Astrakharchik2008,Citro,Silvi:13}. 

For sufficiently large frequencies $\omega_t$ (or, alternatively, small linear densities $1/a$), we find a single array, or linear configuration, as in Fig. \ref{fig:1}(a). Its equilibrium radius $R$  is larger than the confining radius $R_0$ due to the repulsive interactions.   For $\omega_t<\omega_t^{(c)}$ and a sufficiently large number of particles  the minimal energy configurations determined numerically are inhomogeneous. In particular, they result to be a mixture of single- and two-ring structures, as shown in Fig.~\ref{fig:1}(b). The inhomogeneous configurations  appear when the number of dipoles exceeds a certain value $N_0>32$, and they are thus absent for $N=16$, which was the case reported in Ref. \cite{Astrakharchik2008,Astrakharchik2009}.   
For this parameter range the homogeneous double ring (zigzag configuration) is metastable, separated by a small {\color{black} energy barrier from the linear chain. Both structures are at higher energy than the inhomogeneous one, which exhibits domains of linear and zigzag configurations.}  By further decreasing $\omega_t$ the global minimum is the zigzag configuration, whose equilibrium positions are given by  $r_j=R+(-1)^j b$ and $\phi_j= 2\pi j/N$, where $b>0$ is half the radial distance between the two rings.  
The zigzag configuration is illustrated in Fig. \ref{fig:1}(c). It is found provided the number of particles is even, while for odd $N$  the structure exhibits topological defects \cite{Cartarius}.  

\begin{figure}
 \subfloat{\includegraphics[width=0.24\textwidth,valign=t]{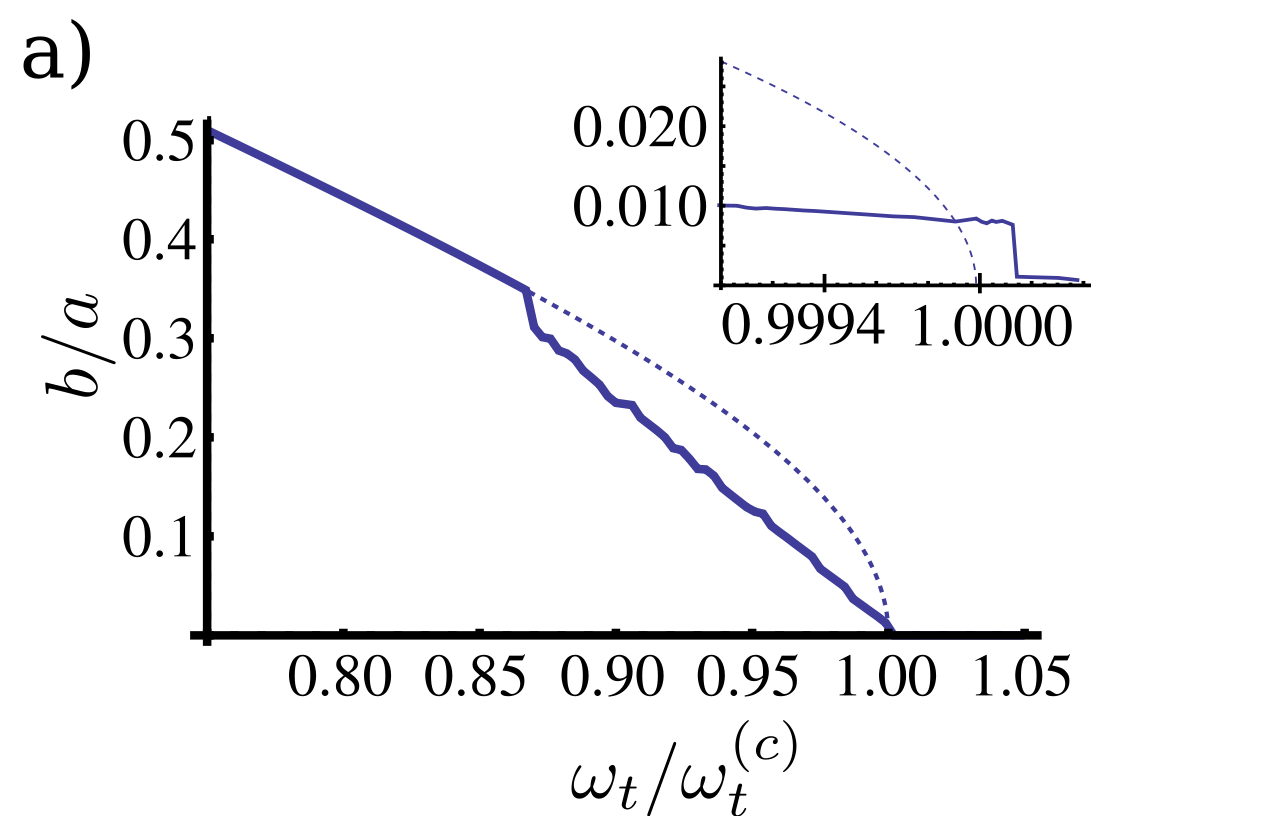}}
\subfloat{ \includegraphics[width=0.24\textwidth,valign=t]{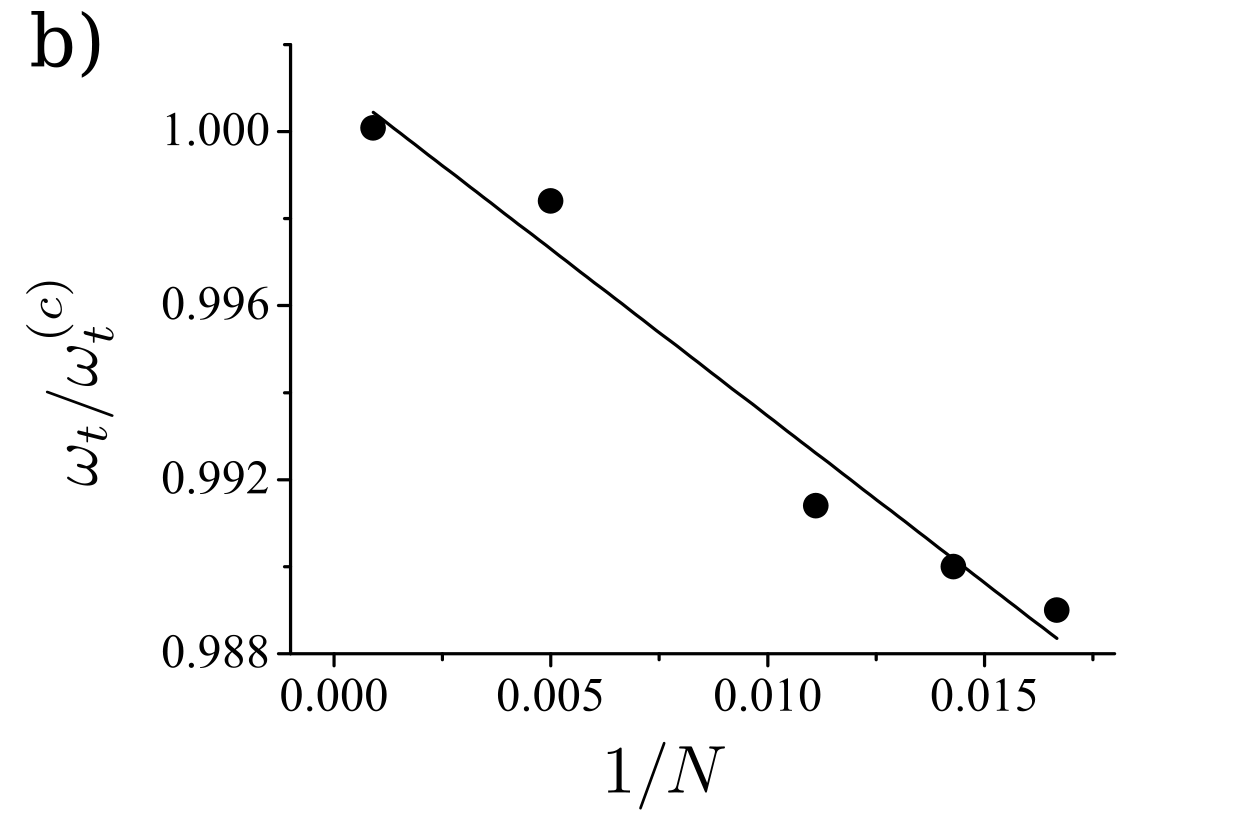}}
 \caption{(Color online) (a) Average transverse displacement $b$ (solid line) along the ring (in units of the interparticle distance along the ring) as a function of $\omega_t/\omega_t^{(c)}$ for N=500 dipoles.
 The dashed line indicates the average displacement of a continuous transition, that is obtained by only allowing transverse particle movement.
 The inset shows the displacement $b$ close to the transition region for 1100 dipoles.
 (b) Trap frequency below which inhomogeneous configurations are the minimal energy solutions in the numerical simulations as a function of 1/$N$, where $N$ is the number of particles along the ring. The red line is a linear fit $\omega_t/\omega_t^{(c)}=a+b/N$ with parameters $a = 1.0011\pm 0.0009$ and $b=-0.77\pm 0.08$.
\label{fig:displacement}}
\end{figure}

Figure \ref{fig:displacement} displays the average transverse displacement as a function of the trapping frequency as obtained from the Monte-Carlo calculations. The region of inhomogeneous configurations is clearly visible as a deviation from the expected square-root behaviour predicted by the Landau theory for a second-order phase transition \cite{Fishman:08,Astrakharchik2008}. A zoom on the transition region also illustrates how the actual transition occurs quite suddenly (within the numerical accuracy) and at a frequency which is slightly larger than the frequency $\omega_t^{(c)}$. The  frequency $\omega_t$ below which inhomogeneous configurations are found tends asymptotically to the value $\omega_t=1.0011(9)\omega_t^{(c)}$. Finite-size corrections scale linearly with $1/N$, as illustrated in Fig.\ref{fig:displacement} (b). 

 The results presented here are {\it not} a peculiarity of the ring geometry and of the power-law exponent $\alpha=3$.  We have also run Monte-Carlo simulations  for linear traps with hard walls as boundaries, and for particles on a ring with  other power-law interactions with $\alpha>2$. In both cases we have found inhomogeneous configurations, similar to those reported here.  For Coulomb interactions, on the other hand, we have found a homogeneous ground-state solution, in agreement with the results of Ref.~\cite{Fishman:08}. In the Coulomb case, indeed, the inhomogeneous configurations are excitations \cite{Landa:NJP}, and the linear-zigzag transition is continuous \cite{Fishman:08}. Our numerical results clearly indicate that the structural transition for dipolar gases (and in general for $\alpha>2$) deviates from the behaviour predicted from the Landau theory for second-order phase transitions.  

\section{Analysis of the structural transition}
\label{res:low-energy}

Since at the mechanical instability second-order derivatives \cs of the potential energy vanish, the thermodynamic properties in this parameter region can be analytically determined by considering higher-order terms in the Taylor expansion.
For this purpose  we derive here an expression of the potential-energy functional at low energies. This then allows us to gain analytical insight of the numerical results.

\subsection{Low-energy model}

To proceed, we recall that close to the structural transition low-energy excitations correspond to normal modes in the longitudinal (tangential) direction with wave numbers $|k|a\ll 1$, and in the transverse (radial) direction with $|k-k_0|a\ll 1$. The latter are long-wavelength excitations of the staggered field $\Psi_{j,st}=(-1)^j\Psi_j$.  The procedure  \cs is a straightfoward extension 
of the one performed  for Coulomb interactions  in Ref.~\cite{Fishman:08,DeChiara:10}, to which we refer for further details of the derivation. 
Keeping just the modes within this low energy cutoff and going back to real space, one can resort to a continuum theory, introducing now the fields as a function of the continuous variable $x$:
\begin{eqnarray}
\Psi(x) \rightarrow & \frac{1}{\sqrt{N}} \sum_k \tilde{ \Psi}_k e^{i k x a} \, ,\\
\Theta(x) \rightarrow & \frac{1}{\sqrt{N}} \sum_k \tilde{ \Theta}_k e^{i k x a} \, ,
\end{eqnarray}
where the coordinate $x$ is in units of the average interparticle distance $a$. With this low-energy cutoff one obtains an expression for the potential energy,  $V_0 = V^{\text{eq}} + V_0$, where $V^{{\color{black}\text{eq}}}$ is the equilibrium energy of the single ring and
\begin{align}
V_0\!&=\!\! \frac{C_D}{a^\alpha} \!\!\!  \int \!\!\! \dd x 
  \left[  h_1^2 (\partial_x\Theta)^2 \!\!+ h_2^2 (\partial_x\Psi)^2 \!\!+ \Delta \Psi^2  \!\! + e(\partial_x\Theta) \Psi^2  + f \Psi^4  
  \right. \nonumber \\
  &\left.+
  r  (\partial_x\Psi )^2 \Psi^2+ \ell (\partial_x\Theta) ^2 \Psi^2 
 + t \Psi^6 + p (\partial_x\Theta)^3 + q \Psi^4 \partial_x\Theta \right] \, , \label{eq:cont-en}
\end{align}
and all parameters are dimensionless constants defined in Appendix \ref{App:A}. Expression \eqref{eq:cont-en} differs from the one reported in Ref. \cite{DeChiara:10} since it contains an expansion up to 6th order as well as the coupling between axial and transverse modes. For Coulomb repulsion this coupling leads to a renormalization of the coefficients, such that sufficiently close to the zigzag instability one can reduce the potential to an effective $\phi^4$ model and neglect higher order corrections.  The inhomogeneous configuration found numerically, however, suggest that for $\alpha>2$ this coupling may play a relevant role. 

\subsection{Minimum energy configurations}
\label{thermo-limit}

\begin{figure}
 \subfloat{\includegraphics[width=0.24\textwidth]{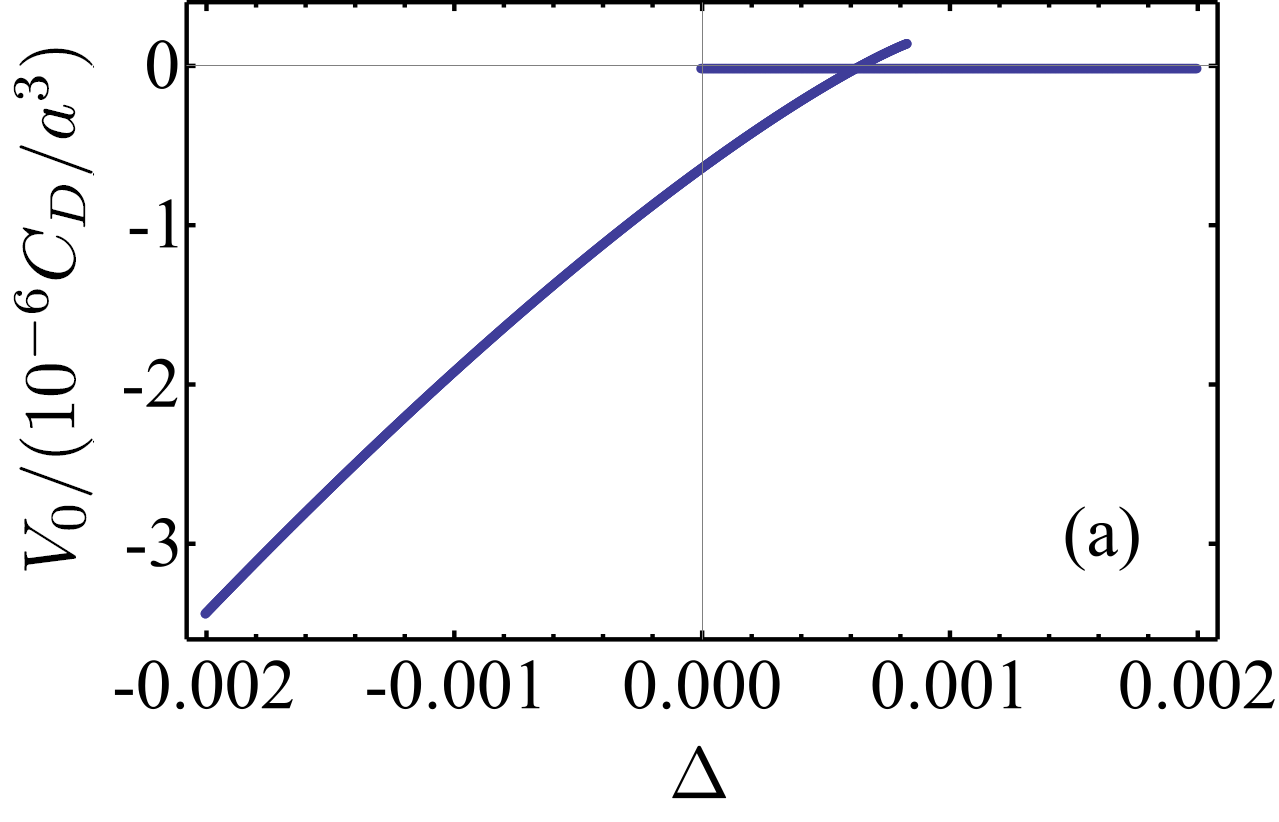}} %
 \subfloat{\includegraphics[width=0.24\textwidth]{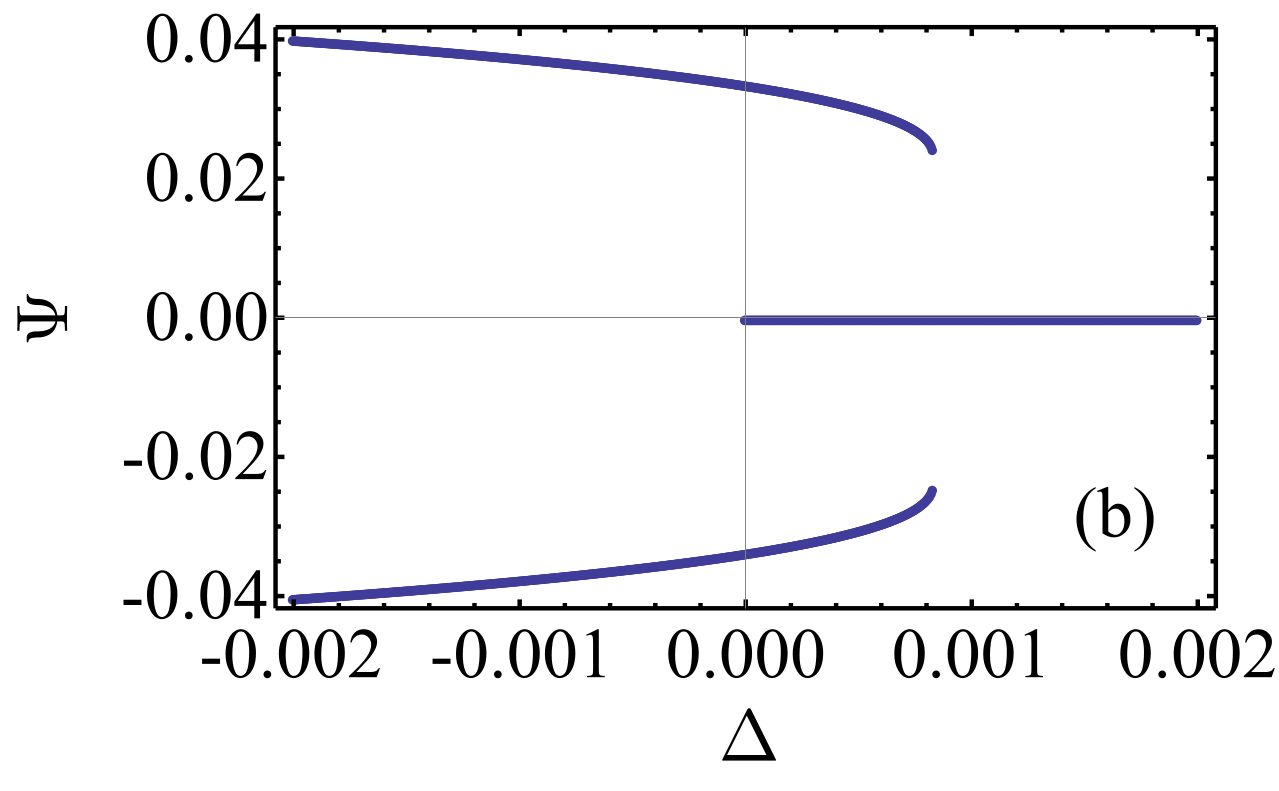}}
 \caption{(Color online) (a) Local minima of the energy in Eq.~(\ref{eq:cont-en}) for homogeneous solutions ($\Psi' = 0$) and (b) corresponding transverse-displacement field (in units of distance $a$ along the chain) as a function of $\Delta$ (dimensionless) and in the thermodynamic limit. The region of coexistence of phases is in the interval $\Delta\in [0,0.0009]$, corresponding to $\omega_t^{(c)}\le\omega_t \le 1.000075 \ \omega_t^{(c)}$.
\label{fig:psi6}}
\end{figure}

 In order to get an insight into the nature of the transition,
we now look for uniform solutions for the fields $\Psi$ and $\Theta'=\partial_x\Theta$  minimizing the long-wavelength potential energy  (\ref{eq:cont-en}) for different values of $\Delta$, and thus of $\omega_t$.
This allows us to find an analytical solution, with which we can verify whether there exists a parameter regime where the linear and the zigzag configurations are {\color{black} both local minima of the potential energy.}
The solutions are extrema of the potential, satisfying $\partial V_0/\partial \Theta' = 0$ and $\partial V_0/\partial \Psi = 0$ with positive-definite  Hessian matrix. 
We determine an effective potential for the transverse-displacement field $\Psi$ by eliminating the solution  for $\Theta'$, which  in the small-$\Psi$ limit reads
\begin{align}
\label{Thetaprime}
\Theta'=- \frac{1}{2 h_1^2} \Psi^2 \left[e +\left(q - \frac{e l}{h_1^2} + \frac{3 e^2 p}{4 h_1^4} \right)\Psi^2\right].
\end{align}
Note that there is a second solution for $\Theta'$, which is finite at small $\Psi$, and thus inconsistent with our initial assumptions.  Substitution of Eq. \eqref{Thetaprime} in the expression (\ref{eq:cont-en}) leads to the effective potential density
\begin{equation}
V_{\rm eff}\propto \Delta \Psi^2 + u_{\rm eff} \Psi^4/4 + \lambda \Psi^6 \, ,
\end{equation}
where $u_{\rm eff}=(4 f-e^2/h_1^2)$ and $\lambda= \left( \frac{l e^2}{4 h_1^4} - \frac{e^3 p}{8 h_1^{6}} - \frac{e q}{2 h_1^2} + t \right)$. Using the explicit form of the coefficients \cs for the case of dipolar interactions  (see App. \ref{App:A}) we obtain that $u_{\rm eff}<0$ and $\lambda>0$. The effective model thus describes a first-order phase transition at $\Delta=0$. It is interesting to point out that the sign of the quartic term is negative due to the coupling with the axial vibrations. Figure~\ref{fig:psi6} shows the energy of the local minima and the corresponding displacement field  $\Psi$ obtained from the low-energy effective model as a function of the control parameter~$\Delta$.

This solution predicts a sudden jump into two stable local minima near the dynamical instability of the single ring, which is  characteristic of a first-order transition. {\color{black} Note that this solution is restricted to uniform transverse fields. Numerically, we find that the inhomogeneous solution is at lower energy, corresponding to  the coexistence of the  zigzag and linear configurations}. Quite remarkably, the parameter region of coexistence of phases is very narrow and close to the frequency $\omega_t^{(c)}$. Therefore, this transition is of 'weakly first-order' or of nearly second order \cite{Larkin,Imry}.

\subsection{Finite-size system}

We now address the predictions of the low-energy model for the displacement fields $\Theta$ and $\Psi$ in a ring of finite size. An analytical solution can be obtained if we keep just the leading order in the transverse-axial coupling, after setting $r,\ell,t,p,q=0$ in Eq. \eqref{eq:cont-en}. This corresponds to a truncation of the effective potential to fourth order. This approach is clearly not capable to describe the nature of the phase in the thermodynamic limit, \cs since it misses the sixth-order terms which stabilize the uniform solution. Nevertheless, in the finite-size ring,  the solution is inhomogeneous, stabilized by the presence of the gradient terms in \eqref{eq:cont-en} and can be employed  to account for the observed inhomogeneous configurations close to the transition point.

Using the variational principle we determine the equations for ~$\Psi(x)$ and $\Theta(x)$ which minimize Eq.\eqref{eq:cont-en},
\begin{align}
 \frac{\dd }{\dd x} \left( 2 h_1^2 (\Theta') + e \Psi^2 \right) =& 0 \label{eq:el-theta} \, , \\
 2 h_2^2 \Psi '' - 2 \Delta \Psi - 2 e \Theta ' \Psi - 4 f \Psi^3 =& 0 \, . \label{eq:el-psi}
\end{align}
These equations admit
  an inhomogeneous soliton-like solution, of the form \cite{Carr,Cominotti}
\begin{align}
  \Psi^2(x) =& y_3 \, \text{cn}^2\left( \frac{\sqrt{g (y_3-y_1)}}{2} x \Big| m \right)\,, \label{eq:soliton} \\
  \Theta '(x) =& \frac{1}{2} B  - \frac{1}{2} \frac{e}{h_1^2} \Psi^2(x)\label{Theta}\,,
\end{align}
where cn is a Jacobi elliptic function and $y_1$, $y_3$, and $B$ are determined by solving coupled transcendental equations, while $m=y_3/(y_3-y_1)$ and $g=-u_{\rm eff} /h_2^2$ (see Appendix \ref{App:B}).

\begin{figure}
\subfloat{
 \includegraphics[width=0.24\textwidth]{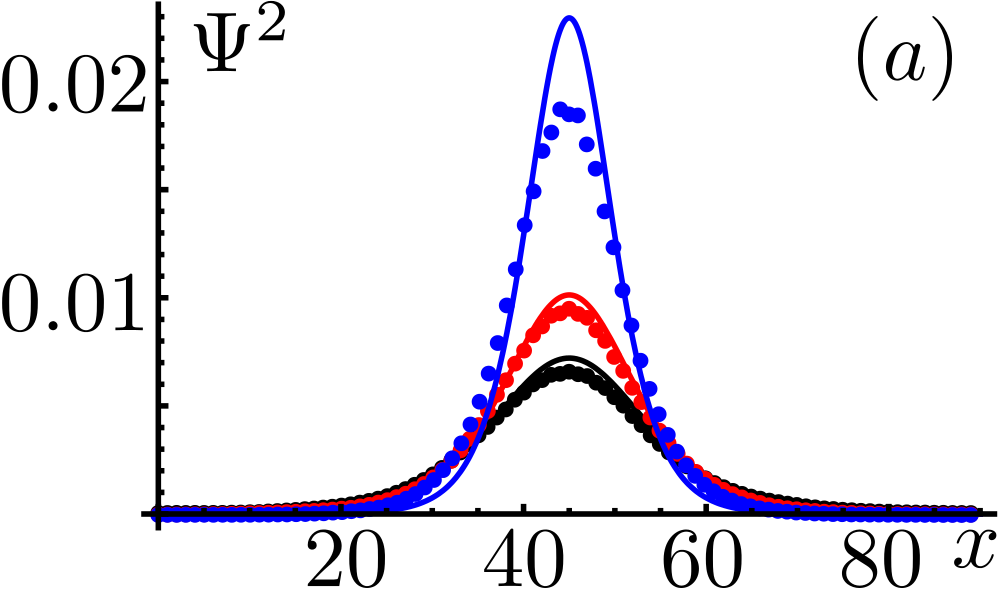}}%
\subfloat{
\includegraphics[width=0.24\textwidth]{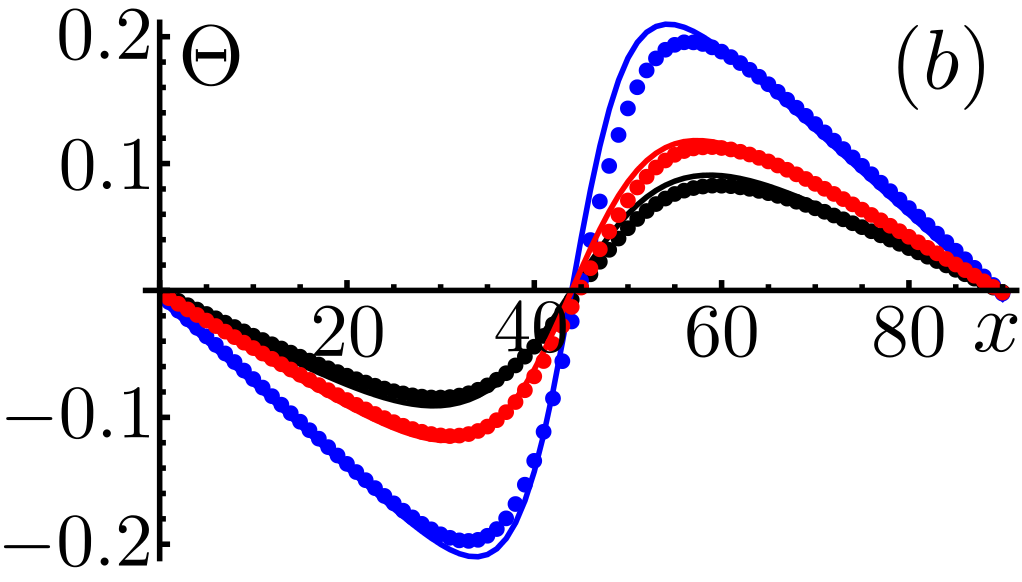}}
 \caption{(Color online) (a) Transverse squared displacement $\Psi^2(x)$ and (b) axial displacement $\Theta(x)$ (in units of  distance $a$ along the chain) as a function of distance $x$ along the chain (in units of $a$) for the minimal energy configurations on  a ring with $N=90$ particles. Numerical Monte-Carlo data (circles)  are compared to the solutions of Eqs.~\eqref{eq:soliton}-\eqref{Theta} (solid lines). From top to bottom, the blue, red and black curves correspond to trap frequencies  $\omega_t= 0.9915 \omega_t^{(c)}$, $0.99 \omega_t^{(c)}$, $0.985 \omega_t^{(c)}$, respectively.\label{Fig:soliton}}
\end{figure}

Figure~\ref{Fig:soliton} displays the behaviour predicted by Eqs. (\ref{eq:soliton}-\ref{Theta})  along the chain and the corresponding numerical results, showing a very good agreement within the model's regime of validity.  The energy of the inhomogeneous configurations is obtained by substituting the corresponding solutions into the potential-energy density. It  is found to be smaller than the energy of the zigzag case, in full agreement with the numerical observations. 
Inspection of Fig.~\ref{fig:displacement} shows that in the numerical calculations for a finite ring the parameter region of phase coexistence is larger than in the thermodynamic limit, extending to negative values of $\Delta$.  This can be explained noticing that  boundary effects yield a renormalized control parameter  $\Delta_{\rm eff}$ for the transition. Details are reported in Appendix \ref{App:B}.

\section{Discussion and conclusions}
\label{Sec:5}

\begin{figure}[t]
\includegraphics[width=0.5\textwidth]{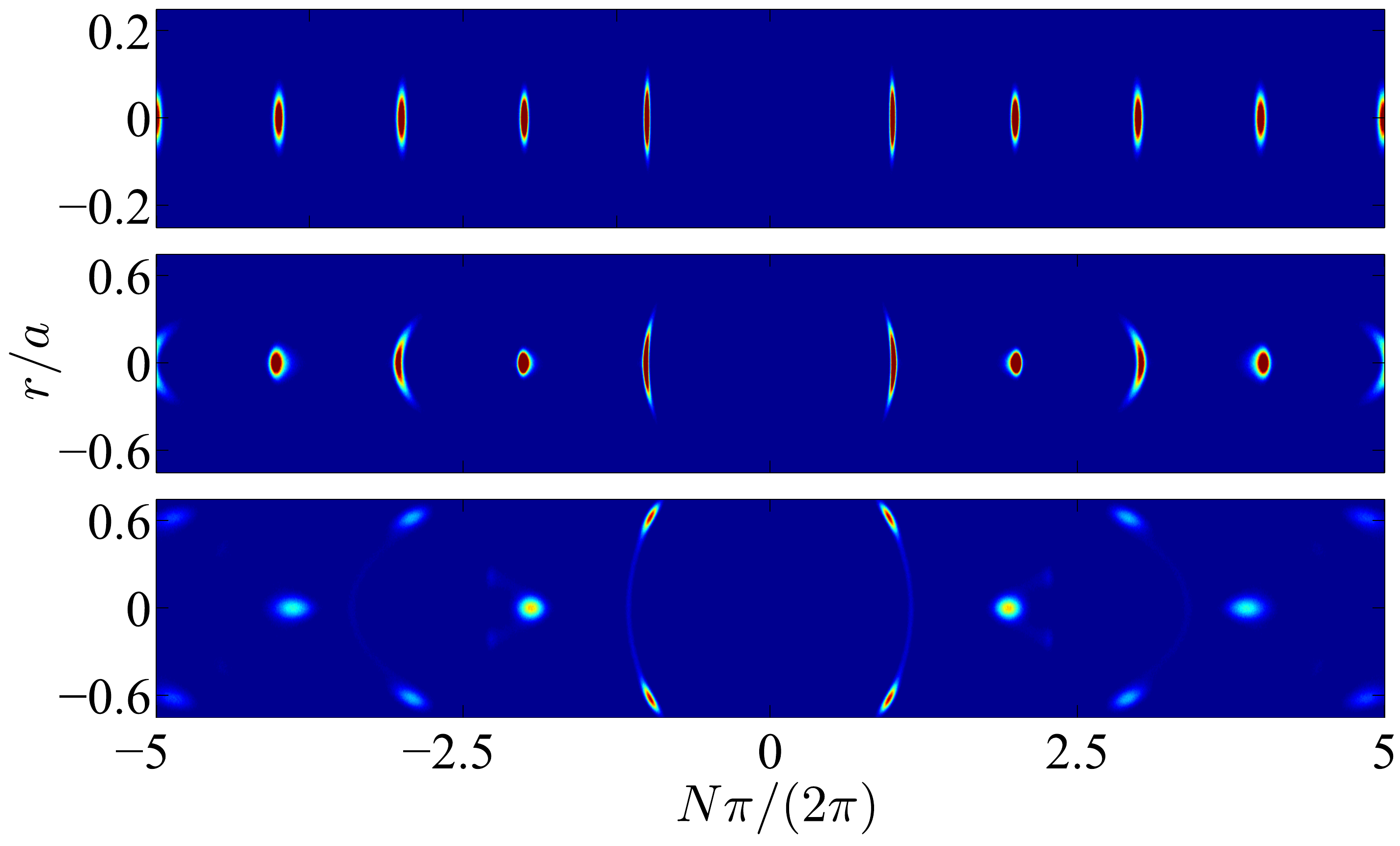}
 \caption{(Color online) Two-particle correlator $g_2(r,\phi)$ of 90 dipoles evaluated numerically and at $T=8 \times 10^{-4} \ C_D/ (a^3 k_B)$. The configurations correspond to a uniform single-ring (top, $\omega_t=1.05 \omega_t^{(c)}$),  an inhomogeneous structure (center, $\omega_t=0.98 \omega_t^{(c)}$) and uniform two-ring configuration (bottom,  $\omega_t=0.7 \omega_t^{(c)}$).
 \label{fig:g2}}
\end{figure}

Our predictions are strictly valid when the effect of fluctuations is negligible. To study the effect of thermal fluctuations on the various configurations found at zero temperature, we have performed a finite temperature Monte-Carlo calculation, and determined the  pair correlation function $g_2(r,\phi) = \langle \sum_{i,j\neq i} \delta (r - (r_i-r_j)) \delta (\phi - (\phi_i-\phi_j)) \rangle$ for temperatures which are lower than the difference between the inhomogeneus and zigzag energies. Figure \ref{fig:g2} displays the two-particle correlation functions for different values of $\Delta<0$. The inhomogeneous configurations are clearly visible as the correlation is smeared along the radial direction in a semicircular shape, indicating varying radial displacements (thus, inhomogeneous $\Psi(x)$). This result for the pair-correlation function is considerably different from both the  one for the linear configuration, characterized by a periodic structure only along the tangential (axial) direction, and  the one for a uniform two-ring configuration, where radially the only possible relative distances allowed are $\pm \Psi$ and 0. The  clear distinction between the various configurations is lost for temperatures higher than the energy {\color{black} barrier} between the various configurations. Taking the value of the dipolar moment of  LiCs molecules~\cite{LiCs} and typical densities of the ongoing experiments \cite{Wu:2012},  we estimate that the energy gap between the  {\color{black} inhomogeneous and uniform configurations} corresponds to a temperature of 0.2 nK. Although this value is  still quite challenging from an experimental point of view, it can rapidly increase at increasing the density and the dipolar moment of the gases. 

To estimate the parameter range for which the system is in a classical regime, we can compare the length scale associated with the quantum fluctuations $a$, with the length scale associated with the interactions $r_0$, which can be estimated to be $r_0 = m C_D / \hbar^2$~\cite{Astrakharchik2008}. If $a \ll r_0$, the ground state energy of the system is well approximated by the classical ground state energy. In this regime, the quantum fluctuations have a similar effect as the temperature has in a classical system~\cite{Astrakharchik2009}. For LiCs molecules, the characteristic length is given by $r_0 = 63 \, \mu$m. Taking a Gaussian wave packet of the same size, the kinetic energy of a molecule can be estimated to be $E \approx k_B \cdot 9 \, \mu$K, which is larger than the energy gap of 0.2 nK. Thus, for the parameters of LiCs molecular gases, it is expected that quantum fluctuations will smear the transition.

In conclusion, we have shown that the linear-zigzag instability for power-law interactions $\alpha>2$ is a first-order phase transition, even though weak, whose hallmark is the appearance of inhomogeneous soliton-like structures which minimize the energy of finite systems. The instability is thus not described by a $\phi^4$ model, since the coupling with the axial vibrations substantially modifies the properties of the transition. This is different from Coulomb systems, where the dispersion relation of the axial modes leads just to a renormalization of the coefficient of the $\phi^4$ model in the critical region, without changing its nature \cite{Silvi:unpublished}. The dipolar system therefore realizes an example of Ising model coupled to axial phonons  \cite{Larkin,Imry}.
 Whether the weakly first-order nature of the transition survives the inclusion of quantum fluctuations is a question for future work. 
In the quantum regime, the instability is expected to exhibit the existence of a critical point with enhanced symmetry and nonuniversal critical exponents, in analogy to the model discussed in Ref. \cite{Sitte:2009}.

\begin{acknowledgments}
The authors are grateful to Eugene Demler, Shmuel Fishman, Frank Hekking, Julia Meyer, Efrat Shimshoni, and Pietro Silvi for stimulating discussions and helpful comments. GM acknowledges hospitality by the ion storage group at NIST, Boulder, during completion of this work. Financial support from the European Commission (STREP PICC), the German Research Foundation, the Handy-Q ERC grant N.\;258608 and the ANR  project no. ANR-13-JS01-0005-01 is acknowledged. 
\end{acknowledgments}

\onecolumngrid
\appendix

\section{Definitions of the expansion coefficients}
\label{App:A}

The coefficients of the potential in Eq. \eqref{eq:cont-en} are given by

\begingroup
\allowdisplaybreaks
\begin{align}
 h_1^2 =& \frac{1}{4} \sum_{l \neq 0}l^2 \frac{a^2}{R^2}\frac{\partial^2 \tilde U(l)}{\partial \phi^2} \label{defh1} \\
 h_2^2 =& -\frac{1}{4} \sum_{l \neq 0} (-1)^l (la)^2 \left( \frac{\partial^2  \tilde U(l)}{\partial \rho^2} - \frac 1 4 \frac{\partial^2  \tilde U(l)}{\partial R^2} \right) \label{defh2}
\\
 \Delta =& (\omega_t^2 - \omega_t^{(c)2}) \,\frac{m a^{\alpha + 2}}{C_D}  \label{defdelta} \\
 e =& \sum_{l \neq 0} l^2 \frac{a^3}{R} \left( \frac 1 4  \cos^2\left(\frac{l \pi}{2}\right) \frac{\partial^3 \tilde U(l)}{\partial R^2 \partial \phi} - \sin^2\left(\frac{l \pi}{2}\right) \frac{\partial^3 \tilde U(l)}{\partial \rho^2 \partial \phi} \right)  \label{defe}
 \\
 f =& \frac{1}{3} \sum_{l \neq 0} a^4\left( \frac{\partial^4 \tilde U(l)}{\partial \rho^4} \sin^4\left(\frac{l \pi}{2}\right) + \frac{1}{16} \frac{\partial^4 \tilde U(l)}{\partial R^4} \cos^4\left(\frac{l \pi}{2}\right) \right)  \label{deff}\\
 \ell =& \sum_{l\neq 0} l^2 \frac{a^4}{R^2} \left( \frac{1}{8} \ \cos^2\left(\frac{l \pi}{2} \right) \frac{\partial^4 \tilde U(l)}{\partial R^2 \partial \phi^2} - \frac{1}{2}  \sin^2\left(\frac{l \pi}{2} \right) \frac{\partial^4 \tilde U(l)}{\partial \rho^2 \partial \phi^2}  \right) \\
 r =& \sum_{l\neq 0} l^2 a^4 \left( \frac{1}{2} \sin^4\left(\frac{l \pi}{2} \right) \frac{\partial^4 \tilde U(l)}{\partial \rho^4}
 - \frac{1}{32} \cos^4\left(\frac{l \pi}{2} \right) \frac{\partial^4 \tilde U(l)}{\partial R^4} 
 + \frac{1}{8}  \frac{\partial^4 \tilde U(l)}{\partial R^2 \partial \rho^2}  \right) \\
 t =& \sum_{l \neq 0}a^6 \left( -\frac{2}{45} \sin^6\left( \frac{l \pi}{2} \right) \frac{\partial^6 \tilde U(l)}{\partial \rho^6}
 + \frac{1}{1440} \cos^6\left( \frac{l \pi}{2} \right) \frac{\partial^6 \tilde U(l)}{\partial R^6} 
  \right) \\
 p =& \frac{1}{12} \sum_{l \neq 0} l^3 \frac{a^3}{R^3} \frac{\partial^3 \tilde U(l)}{\partial \phi^3}\\
 q =& \sum_{l \neq 0} l \frac{a^5}{R} \left( -\frac{1}{3}  \sin^4\left(\frac{\pi l}{2} \right) \frac{\partial^5 \tilde U(l)}{\partial \rho^4 \partial \phi} - \frac{1}{48}  \cos^4\left(\frac{\pi l}{2} \right) \frac{\partial^5 \tilde U(l)}{\partial R^4 \partial \phi}\right)\,,
\label{defq}
\end{align}%
\endgroup
where we introduced $\tilde{U}=U/(C_D/(a^\alpha))$.\\

\twocolumngrid

\section{Soliton solutions}
\label{App:B}

In order to obtain the solutions of Eqs.(\ref{eq:el-theta}-\ref{eq:el-psi}),  
we start by integrating Eq.~(\ref{eq:el-theta}), obtaining
\begin{align}
 \Theta '  = \frac{1}{2} B - \frac{1}{2} \frac{e}{h_1^2} \Psi^2, \label{eq:el-theta2}
\end{align}
where $B$ is an integration constant.
Substituting Eq.~(\ref{eq:el-theta2}) into Eq.~(\ref{eq:el-psi}) gives
\begin{align}
-2 h_2^2 \Psi ''  + 2 \Delta_{\rm eff}  \, \Psi + u_{\rm eff}   \, \Psi^3=0, \label{eq:el-psi2}
\end{align}
where $\Delta_{\rm eff} = \Delta + e  B/2 $ and $u_{\rm eff}= 4 f - e^2/h_1^2$ are the renormalized constants entering the resulting effective potential-energy functional 
 $V_{\rm eff}=\frac{C_D}{a^\alpha}  \int \dd x 
  \left[ h_2^2 (\Psi')^2 + \Delta_{\rm eff} \, \Psi^2  + \frac{u_{\rm eff}}{4} \, \Psi^4 \right]$.
Note that in the finite ring the boundary conditions effects yield a renormalization of the constant $\Delta$.  This explains why the region of phase coexistence extends to negative values of $\Delta$ for finite systems (see Fig.~\ref{fig:displacement}).
Multiplying Eq.~(\ref{eq:el-psi2}) by $\Psi'$ and a subsequent integration leads to
\begin{align}
 (\Psi ') ^2 = \frac{1}{h_2^2} \Delta_{\rm eff}   \, \Psi^2 +  \frac{1}{4 h_2^2} u_{\rm eff}  \, \Psi^4 + \frac{1}{4} A,
\end{align}
where $A$ is another integration constant. As this equation only depends on $\Psi^2$, we perform the substitution~$y = \Psi^2$ and obtain
\begin{align}
 (y')^2 = \frac{4}{h_2^2}  \Delta_{\rm eff}   \, y^2 + \frac{1}{h_2^2} u_{\rm eff} \, y^3 + A y. \label{eq:el-psi4}
\end{align}
This equation can be solved by separating the variables \cite{Carr}. We define the zeros of the right hand side of Eq.(\ref{eq:el-psi4}) as $y_1 < y_2 < y_3$ and set $ g = - u_{\rm eff}/4 h_2^2$. Eq.~(\ref{eq:el-psi4}) can be integrated as
\begin{align}
 \int_0^x \dd \tilde x = \int_{y_3}^{y} \frac{\dd \tilde y}{\sqrt{-g (\tilde y-y_1)(\tilde y-y_2)(\tilde y-y_3)}}.
\end{align}
Finally we perform the substitution $t^2 = \frac{\tilde y-y_2}{y_3-y_2}$ and with 
\begin{align}
 m = \frac{y_3-y_2}{y_3-y_1} = 1-m' ,
\end{align}
we arrive at
\begin{align}
 \int_0^x \dd \tilde x = \frac{2}{\sqrt{g(y_3-y_1)}} \int^{Y}_{1} \frac{\dd t}{\sqrt{(1-t^2)(m t^2 + m')}}\,,
\end{align}
where $Y=\sqrt{(y-y_2)/(y_3-y_2)}$. This equation can be solved as
\begin{align}
 y(x) = \Psi^2(x) = y_3 \,\text{cn}^2\left( \frac{\sqrt{g (y_3-y_1)}}{2}x | m \right), \label{eq:app-soliton}
\end{align}
where $\text{cn}(x|m)$ is a Jacobi elliptic function.
The soliton discussed here is given by the case $y_2=0$. As our system is periodic, we will shift $x$ by $N/2$, to center it between $0$ and $N$.
The remaining constants $y_1$ and $y_3$ depend on the constants in the potential energy density in Eq.~(\ref{eq:cont-en}) and the integration constants~$A$ and $B$, which are determined by the boundary conditions,
\begin{align}
 y(0) =& y(N), \\
 \Theta(0) =& \Theta(N).
\end{align}
Combining both boundary conditions, we find
\begin{align}
 \frac{2 K(m)}{N} =& \frac{\sqrt{g(y_3-y_1)}}{2}, \label{eq:boundary1} \\
 B =& 16 \frac{e}{g h_1^2} \frac{1}{N^2} K(m) \left( E(m) + (m-1)K(m) \right), \label{eq:boundary2}
\end{align}
where $K(m)$ and $E(m)$ are the complete elliptic integrals of the first and second kind, respectively and by solving eqs.~(\ref{eq:boundary1}) and (\ref{eq:boundary2}), the two integration constants can be determined.
By substituting eq.~(\ref{eq:app-soliton}) into the long wavelength potential energy  we finally determine the energy of the soliton solution.

\bibliographystyle{prsty}

\end{document}